\documentstyle[11pt,aas2pp4]{article}
\def \lya {Ly$\alpha$}
\def \lyb {Ly$\beta$}
\def \ha {H$\alpha$}
\def \hb {H$\beta$}
\def \oIII {[O~III]~$\lambda\lambda 4959, 5007$}
\def \kms {~km~s$^{-1}$}
\def \cmII {cm$^{-2}$}
\def \cmIII {cm$^{-3}$}
\def \flux {~erg~s$^{-1}$~cm$^{-2}$~\AA$^{-1}$}
\def \zw {I~Zw~1}

\begin{document}

\title{THE UV PROPERTIES OF THE NARROW LINE QUASAR I~Zwicky~1}
\author {Ari Laor}
\affil{Physics Department, Technion, Haifa 32000, Israel}
\authoremail{laor@physics.technion.ac.il}
\author{Buell T. Jannuzi, Richard F. Green, and Todd A. Boroson} 
\affil{National Optical Astronomy Observatories, P.O. Box 26732, 
950 N. Cherry, Tucson, AZ 85726-6732}

\begin{abstract}

I~Zwicky~1 (\zw) is the prototype narrow line quasar. Its narrow line
profiles minimize blending effects and thus allow a significantly more
accurate study of individual quasar emission lines.  We report here
the results of our study of the 1150-3250~\AA\ emission of \zw\ using
a high S/N (50-120) spectrum obtained with the Faint Object
Spectrograph (FOS) of the {\it Hubble Space Telescope} (HST). The UV
spectrum of \zw\ is very rich in emission lines, with essentially no
wavelength range free of emission line features. The following main
new results are obtained:

\begin{enumerate}

\item The Mg~II~$\lambda 2798$ doublet is partially resolved, and
the measured doublet ratio is 1.2/1. This ratio verifies theoretical
predictions that the Mg~II~$\lambda 2798$ line is thermalized 
in the Broad Line Region (BLR).

\item The Al~III~$\lambda 1857$ doublet is clearly resolved, with
a thermalized doublet ratio of 1.25/1. The line
optical depth provides an upper limit to the clouds distance from
the ionizing source which is
consistent with the ``standard'' BLR radius. 

\item A weak associated UV absorption system is detected in N~V (possibly
also in C~IV and \lya\ as well), suggesting an outflow with a velocity of 
$\sim 1870$\kms\ and velocity dispersion $\lesssim 300$\kms. 

\item Lines from ions of increasing ionization level show increasing
excess blue wing flux, and an increasing line peak velocity shift,
reaching a maximum blueshift of $\sim 2000$\kms\ for He~II~$\lambda 1640$.
This may indicate an out-flowing component in the BLR, 
where the ionization level increases with velocity, and which is
visible only in the approaching direction.
The highest velocity part of this outflow may produce the associated UV
absorption system.

\item The small C III]$~\lambda 1909$ equivalent width (EW), and the small
C III]$~\lambda 1909/$\lya\ and C III]$~\lambda 1909/$Si III]$~\lambda
1892$ flux ratios indicate a typical BLR density of $n_e\sim
10^{11}$~\cmIII, i.e.  about an order of magnitude larger than implied
by C~III]$~\lambda 1909$ in most quasars. A BLR component with a
density $n_e\gtrsim 10^{11}$~\cmIII\ is implied by the EW and doublet
ratio of the Al~III~$\lambda 1857$ doublet.

\item Prominent Fe~II~UV~191 emission is seen, together with weaker line
emission at 1294~\AA\ and 1871~\AA. These three features have been proposed
by Johansson \& Jordan (1984) as evidence for significant \lya\ pumping
of the 8-10 eV levels of Fe~II.

\item Significant Fe~III emission is present. The Fe~III~UV~34 and UV~48
multiplets are clearly resolved, and Fe~III~UV~1, UV~47, UV~50, and
UV~68 may also be present. The implications of significant Fe~III
emission for the conditions in the BLR needs to be explored.

\item Very weak [C III]$~\lambda 1907$ and [Si III]$~\lambda 1883$ emission
may be present, suggesting a Narrow Line Region (NLR) component with 
$n_e\sim 5\times 10^5$~\cmIII.

\end{enumerate}

The rich UV emission spectrum of \zw\ should serve as a useful
``template'' for the identification of weak features in other Active
Galactic Nuclei (AGNs), and as a useful benchmark for photoionization
models, in particular for models of the complex Fe~II emission
spectrum.

\end{abstract}


\section{Introduction}

The prominent broad emission lines of quasars are typically blended
with weaker emission features, such as emission from various Fe~II
multiplets.  This blending prevents a reliable study of the prominent
line profiles, the measurement of weaker lines, and the identification
of very weak lines. Even in high S/N spectra (e.g. Laor et al. 1994a)
the blending of features prevents a reliable study of the individual
line profiles of even the most prominent lines.  Blending effects are
minimized in the rare class of narrow line quasars (i.e. with
FWHM$\lesssim 1000$~km~s$^{-1}$).  Therefore the study of such objects
can lead to a significantly more accurate study of the properties of
individual emission lines.

I~Zwicky~1 (\zw) is the prototype narrow line Seyfert 1 galaxy
(NLS1). Its optical spectrum reveals narrow emission lines and strong
Fe~II emission (Sargent 1968; Phillips 1976, 1978; Oke \& Lauer
1979). The NLS1 class of AGNs also tend to have weak \oIII\ emission,
asymmetric \hb\ profiles (Boroson \& Green 1992, hereafter BG92),
steep soft X-ray spectra (Laor et al. 1994b, 1997a; Boller, Brandt \&
Fink 1996), and possibly strong IR emission (e.g. Lipari 1994).
Although \zw\ is classified as a Seyfert 1 galaxy, its absolute
luminosity, $M_V=-23.8$ (for $H_0=50$\kms~Mpc$^{-1}$, $q_0=0$),
actually qualifies it as a low luminosity quasar (V\'{e}ron-Cety \&
V\'{e}ron 1991), we therefore refer to it as a quasar throughout the
paper.

In this paper we describe a detailed study of the UV emission line
spectrum of \zw\ obtained with the Faint Object Spectrograph (FOS) of
the {\it Hubble Space Telescope} (HST).  An earlier UV spectrum of
\zw\ obtained with the International UV Explorer is described by Wills
(1983), and HST UV spectropolarimetry of \zw\ is described by Smith et
al. (1997).  Our study has two main goals. The first goal is to
provide a detailed ``identification template'' of the UV emission
features present in the spectra of quasars. The second goal of this
study is to provide a high S/N ratio Fe~II UV emission template. The
strong and narrow Fe~II emission lines in \zw\ make it an ideal object
for the construction of such a template, as was clearly demonstrated
in the optical regime by BG92.  Such a template is also important for
studies of UV Fe~II emission in AGNs (e.g. Netzer \& Wills 1983;
Wills, Netzer, \& Wills 1985; Collin-Souffrin \& Dumont 1986).  These
studies are important for understanding the physics of the broad-line
region (BLR) since Fe~II is a major, and sometimes the dominant
coolant of the BLR (e.g. Phillips 1977, 1978; Miley \& Miller 1979;
Bergeron \& Kunth 1984; Joly 1981; Lipari, Terlevich, \& Macchetto
1993).  The Fe~II template is also important for studies of lines
strongly blended with Fe~II, such as Mg~II~$\lambda 2798$.

In this paper we concentrate on the first goal of this project. In \S
2 we describe the observations and present the ``identification
template''.  In \S 3 we discuss various diagnostics provided by some
of the UV lines, and in \S 4 we describe the physical properties of an
associated UV absorption system, and discuss its possible relationship
to observed UV emission and X-ray absorption. A brief discussion of
the similarities of \zw\ to broad absorption line quasars (BALs) (\S
5) and a summary of our main conclusions (\S 6) complete the paper.

\section{Observations and Analysis} 

We obtained high S/N spectra (50$-120$ per resolution element) of \zw\
with the three high resolution (R=1300) gratings of the FOS from
1150~\AA\ to 3280~\AA. This S/N is comparable to the highest S/N yet
obtained with HST for bright AGNs.  All the observations were made
using the 0''.86 aperture.  The dates of the observations are: G130H,
13 February 1994 (UT Date); G190H and G270H, 14 September 1994.  The
HST spectra were reduced as described in Schneider et al. (1993).  The
wavelength zero points of the three HST spectra were set using
Galactic interstellar medium absorption lines (Savage et
al. 1993). The size of the necessary shift and the lines used for each
of the three spectra (G130H, G190H, and G270H) were respectively 0.36
\AA\ (Si II~$\lambda 1526.71$), 1.08 \AA\ (Al II~$\lambda 1670.79$),
and 0.99 \AA\ (Fe~II lines).  We also obtained a complimentary high
resolution ground based spectrum which covers the range from 3183~\AA\
to 4074~\AA.  The spectrum was obtained using the KPNO 2.1m telescope
and Goldcam spectrograph on 22 September 1995 UT. The grating used was
a 837 lines/mm grating, used in 2nd order, providing a dispersion of
approximately 0.47~\AA\ per pixel and a spectral resolution of
1.2~\AA. The Ca~II~$\lambda\lambda 3933.66, 3968.47$ doublet
absorption by the ISM was used to secure a zero point wavelength which
matches the UV zero points, which were also set using lines of low
ionization state ions. The optical spectrum is used in order to get
complete coverage of the 2000-4000~\AA\ region, required for
generating the Fe~II template.

Figure 1 presents the overall optical-UV spectrum of \zw\ from
1100\AA\ to 6000\AA. The spectrum was corrected for the effects of
Galactic extinction of E(B$-$V)=0.1 using the standard Seaton (1979)
reddening curve. The extinction is deduced from the neutral hydrogen
column density of N$_{\rm H~I}=5.07\times 10^{21}$~cm$^{-2}$ taken
from Elvis, Lockman, \& Wilkes 1989), and the relation
E(B$-$V)=N$_{\rm H}$/$4.93\times 10^{21}$ found by Diplas \& Savage
(1994). The rest wavelength was calculated assuming $z=0.0608$, the
redshifts of \ha\ and \hb\ (e.g. Phillips 1976). A composite spectrum
of $\sim 700$ quasars observed from the ground by Francis et
al. (1991, hereafter the Francis et al. composite) is also shown for
comparison. This composite represents well the ``average'' quasar's
continuum, and matches well similar composites constructed for lower
$z$ quasars with HST (see Fig.2 in Laor et al. 1995).

The optical spectrum of \zw\ displayed in Fig.1 was kindly provided by
B. Wills (1996, private communication). It matches the HST spectrum in
the overlap region, 3170-3270\AA, to better than 5\%, and is used here
to demonstrate the overall spectral shape. A flux discontinuity may be
present just blueward of C~IV ($\log \nu =15.29$) since the region at
$\log \nu >15.29$ (obtained with the G190H) was observed 7 months
after the $\log \nu <15.29$ region (obtained with the G130H). The
overall optical-UV power-law slope is $\alpha\approx -1$,
significantly redder than the slope of the typical quasar
($\alpha\approx -0.3$, as displayed by the Francis et al.  composite
or other optically selected samples, e.g.  Neugebauer et
al. 1987). This may indicate that \zw\ is significantly reddened, as
further discussed in \S 3.3. Note that many weak features in the
Francis et al. composite are seen in \zw\ as clear and well resolved
lines.

Figures 2a-2c present an expanded view of the UV spectrum shown in
Fig.1.  The rest frame positions of various lines are marked above the
spectrum.  Lines marked below the spectrum designate various
absorption lines originating in the Galactic interstellar medium
(ISM). All ISM lines detected here are typically seen in FOS spectra
of quasars (Savage et al.  1993). We make no attempt here to identify
the many Fe~II emission blends present. The various Fe~II multiplets
designated in Fig.2 are taken from Moore (1952), and serve only as an
illustration of possible Fe~II features. The number of possible Fe~II
multiplets is extremely large, and a reliable identification requires
detailed theoretical calculations, which are beyond the scope of this
paper.

 Figure 2d presents the optical spectrum. The spectrum was obtained in
a single exposure and is affected by cosmic ray hits, which were
deleted by taking out 5 small patches (width 1-2~\AA) from the
spectrum. The absolute flux calibration is not secure. The wavelength
zero point was set using Ca II~$\lambda\lambda 3934.78, 3969.59$
Galactic interstellar medium absorption lines. Since the HST
wavelength zero point is also based on low ionization species (Si~II,
Al~II, Fe~II), it probably matches well the optical spectrum zero
point.  The HST FOS G270H spectrum which overlaps with the optical
spectrum in the 3183~\AA\ to 3280~\AA\ region is also shown in Fig.2d,
shifted down by a factor of 0.37 in log flux. Apart from the flux
discrepancy, the spectral features match very well. All the optical
Fe~II multiplets from the first 20 optical Fe~II multiplets listed by
Moore (1959) are marked for illustration only (it does appear though
that some of the observed lines agree with expected positions). Note
the absence of the [O~II]~$\lambda 3727$ and [Ne~V]~$\lambda\lambda
3346, 3426$ forbidden lines .

Table 1 lists for each emission feature the measured rest frame
wavelength (defined as the wavelength of the peak flux in the Gaussian
smoothed spectrum), the expected wavelength, the emitting ion, the
observed flux, rest frame equivalent width (EW), full width at half
maximum (FWHM), and the velocity shift of the peak of the emission
line.  These parameters were measured by fitting a set of Gaussians to
each emission feature, as described in Laor et al. (1994a, 1995). In
strongly blended features, or in very weak features, some of the
emission parameters could not be measured reliably, and are therefore
not listed. We preferred not to assign errors to the measured
parameters since such errors may not be very meaningful given the very
complicated overall spectrum and the subjectivity of both the
placement of the continuum and separation of blended features.

Figure 3 presents a comparison of an \ha\ profile to the observed
profile of most of the strong emission features.  Direct comparison of
line profiles is not meaningful in most cases since most lines are
strongly blended (despite their narrow profile in \zw). The comparison
is therefore made using a template profile, which allows us to test if
the observed profile can be reconstructed with a sum of template
components representing the various lines which may contribute to the
blend.  For the template we use the high S/N spectrum of \ha\ obtained
for \zw\ as part of the BG92 study. The \ha\ line is used since it has
the highest S/N, and is least blended. The wavelength used to set the
velocity scale is given in each sub-panel. Most features are blends,
and in each subpanel we list the ratio of fluxes in the blend
components assumed in the fit. When searching for the best fit we
allowed the continuum level to be below the apparent continuum level,
since the observed continuum may be contaminated by weak unresolved
emission (see in particular the Mg~II doublet fit, discussed in \S
3.1.5). Note that the \ha\ profile is slightly asymmetric with some
excess flux on the blue wing (e.g. see the C~IV 1/1 doublet fit). The
template fits were not used for measuring any of the emission line
parameters.

\section{Emission Lines} 

Below we describe some of the BLR diagnostics provided by ratios of lines
from different levels of a given ion, or within certain multiplets of a 
given ion.

\subsection{Line Diagnostics} 

\subsubsection{He}

The $R_{\rm He II}\equiv $He~II~$\lambda 4686$/He~II~$\lambda 1640$
flux ratio can be used as a reddening indicator. Below we explore its
usefulness in \zw.  There is a very weak, if any, feature at the
expected He~II rest wavelength of 1640.4~\AA. The nearest strong
feature is at 1629.5~\AA, which we suggest is highly blueshifted
($\sim 2000$\kms) He~II emission. Alternatively, this feature may be
just another Fe~II feature, many of which are scattered throughout the
spectrum. However, we find the He~II identification more likely since
C~IV and N~V also display large blue shifts ($\sim 900$\kms), and the
He~II velocity shift is typically twice that of C~IV (e.g. Laor et
al. 1995, Table 7). In addition the He~II/\lya\ flux ratio of $\sim
0.03$ measured using the 1629.5~\AA\ feature is typical of quasars
(e.g. Laor et al. 1995, Table 6).

Photoionization models indicate values of about $R_{\rm He II}\simeq
8-11$ (Netzer , Elitzur, \& Ferland 1985).  BG92 did not detect the
He~II~$\lambda 4686$ line. A plausible upper limit on the EW is
0.5~\AA\ using the BG92 high S/N optical spectrum, which implies a
line flux$<5\times 10^{15}$\flux, where the continuum flux level is
obtained from a lower resolution optical spectrum presented in
Fig.1. We thus get $R_{\rm He II}>18$.  However, since the
He~II~$\lambda 1640$ line is very broad and asymmetric, the
He~II~$\lambda 4686$ upper limit may have been somewhat
underestimated, while some of the flux attributed to He~II~$\lambda
1640$ may be due to blended Fe~II features.  Therefore, the $R_{\rm He
II}$ does not provide a useful contraint on the reddening.

\subsubsection{C}

Metal lines are produced mostly by collisional excitation of ions in
their ground level. The C~III$^*~\lambda 1176$ line discussed below is
unique in originating from radiative decay between two relatively high
lying levels. The transition is from the $2p^2\,^3\!P$ level, 17~eV
above the ground level, down to the metastable $2s2p\,^3\!P$ level,
6.5~eV above the ground level (which decays to ground level via the
C~III]$~\lambda 1909$ line).  The typical temperature in the C~III
region in photoionized gas is $<2\times 10^4$~K, and the small
Boltzman factor, $e^{-\chi/kT}<5\times 10^{-5}$ ($\chi=17$~eV),
suggests that the collisional excitation rate would be low and
therefore that the line flux should be small.

The C~III$^*~\lambda 1176$ line is clearly detected in \zw.  The
presence of this line was already suggested by Laor et al. (1995,
Fig.4) in high S/N FOS spectra of PG~1116+215, PG~1216+069, and
PG~1538+477, but these detections were only marginal due to the
effects of absorption by the \lya\ forest, and due to strong blending
with the \lya\ blue wing. These two effects are negligible in \zw, and
the line can therefore be clearly detected. The C~III$^*~\lambda
1176$/\lya\ flux ratio found in those three objects is $0.02-0.05$
versus a comparable ratio of 0.014 found here.  This line was also
detected recently in a composite HST spectrum of quasars at a similar
level (Hamann et al. 1997).

Laor et al. (1997b) briefly discussed various excitation mechanisms
for C~III$^*~\lambda 1176$, and showed that the observed high
C~III$^*~\lambda 1176$/ C~III]~$\lambda 1909$ flux ratio is $\sim 50$
times larger than expected if collisional excitation at typical BLR
conditions is the only significant process.  However, as discussed
below, the BLR density in the C~III region of \zw\ is probably about
an order of magnitude larger than in a typical BLR, which strongly
suppresses C~III]~$\lambda 1909$ emission. In addition, the gas
temperature in the BLR tends to increase with increasing density (due
to thermalization of some of the main coolants), which increases the
C~III$^*~\lambda 1176$ emission, as shown by Korista et al. (1997,
their Figure 3c). The suppression of C~III]~$\lambda 1909$ and
enhancment of C~III$^*~\lambda 1176$ at $n_e\gtrsim 10^{11}$~\cmIII\
can explain the high C~III$^*~\lambda 1176$/C~III]~$\lambda 1909$ flux
ratio in \zw.  Note that dielectronic recombination (Storey 1991) also
contributes significantly to the C~III$^*~\lambda 1176$ flux.  Laor et
al. (1997b) suggested that dielectronic recombination is not
significant in \zw\ based on the C~III$~\lambda 2297$ line flux, but
this suggestion was based on a misquote of Storey (1991), and the
actual values do not rule out a significant contribution from
dielectronic recombination. Resonance scattering of continuum photons
may also contribute significantly to the C~III$^*~\lambda 1176$
emission if the velocity dispersion within the BLR clouds is large
enough (Ferguson, Ferland \& Pradhan 1995; Laor et al. 1997b).

The [C~III]~$\lambda 1907$ line may be present in \zw. This line is
produced by a forbidden transition from the $^3\!P_2$ term in the
$2s2p$ configuration of C~III down to the ground level, with a
critical density of $n_c=5\times 10^5$~\cmIII\ (Osterbrock 1989, Table
3.11). The $^3\!P_2$ term lies 0.007~eV above the $^3\!P_1$ term which
produces the well known C~III]$~\lambda 1909$ line, where $n_c\sim
5\times 10^9$~\cmIII.  Thus, the flux ratio $R_{\rm
C~III}\equiv$[C~III]~$\lambda 1907/$C~III]$~\lambda 1909$ provides a
temperature independent density diagnostic (e.g. Osterbrock 1989,
Fig. 5.5).

The observed C~III]$~\lambda 1909$ line profile appears to be double
peaked with the short wavelength peak at 1906.4~\AA\ matching well the
expected 1906.68~\AA\ position of the [C III] line. This
identification is not secure since excess emission is seen on the blue
side of some of the other emission lines (see \S 3.2), and the
significance of the feature is only marginal. But, an analogous [Si
III]~$\lambda 1883$ line may also be present (see \S 3.1.7), which
implies similar conditions.  If true, then the [C III] line is much
narrower and weaker than the C III] line. The observed $R_{\rm
C~III}\lesssim 0.1$ then implies that some of the C III emission
originates in gas with an electron density $n_e\lesssim 5\times
10^5$~\cmIII.

The only clear detection of the [C~III]$~\lambda 1907$ line in AGNs is
apparent in the high-dispersion IUE spectrum of NGC~4151 displayed by
Bromage et al.  (1985, Fig.1).

The C~III]$~\lambda 1909$ line is very weak in \zw. Both its EW of
$\sim 2.5$~\AA, and the C~III]$~\lambda 1909/$ \lya=0.023 flux ratio
are a factor of 5-10 smaller than typically observed (e.g. Francis et
al. 1991; Laor et al. 1995; Wills et al. 1995).  Since C~III]$~\lambda
1909$ is thermalized at $n_e\gtrsim 5\times 10^9$~\cmIII, the
suppression of C~III]$~\lambda 1909$ may imply that the typical BLR
density in \zw\ is about an order of magnitude larger than indicated
by C~III]$~\lambda 1909$ in typical AGNs, i.e. $n_e\sim
10^{11}$~\cmIII\ instead of $n_e\sim 10^{10}$~\cmIII. Additional
evidence for this interpretation is provided by the Si~III]$~\lambda
1892$ discussed in \S 3.1.7.

\subsubsection{N}

The N~V~$\lambda\lambda 1239, 1243$ doublet components are separated
by the equivalent of $\sim 1000$~\kms, and the narrow line profiles in
\zw\ suggest that this doublet should be resolvable.  However, the N~V
profile shows a large excess flux at the blue wing, and the \ha\
template fit is inadequate (see Fig.3). It does appear, however, that
the doublet flux ratio is significantly lower than the optically thin
2/1 ratio. The physical implication of the observed doublet ratio is
discussed below for the Mg~II and Al~III doublets which are rather
well resolved (\S\S 3.1.5, 3.1.6).

\subsubsection{O}

The O~I~$\lambda 1304$ line is thought to be produced through Bowen
fluorescence. In this process a \lyb\ (1025.72~\AA) photon is absorbed
by O~I in the ground $2p\,^3\!P$ level, and gets excited to the
$3d\,^3\!D$ level, which is within the thermal core of \lyb\
(1025.77~\AA).  The decay back to the ground level is either direct,
or through two intermediate levels. The indirect decay route results
in a $\lambda 11287$ photon ($3d\,^3\!D$ to $3p\,^3\!P$), a
$\lambda8446$ photon ($3p\,^3\!P$ to $3s\,^3\!S$), and a $\lambda
1304$ photon ($3s\,^3\!S$ to $2p\,^3\!P$).  Thus, one expects equal
number of photons in these three lines.  Rudy, Rossano \& Puetter
(1989) very nicely confirmed this prediction for the $\lambda 11287$
and $\lambda 8446$ lines in \zw.  Below we show that the same number
of photons is not seen in the $\lambda 1304$ line, and briefly discuss
possible implications.

The O~I~$\lambda 1304$ line is blended with the Si~II~$\lambda\lambda$
1304, 1309 doublet. This blending is very strong in most quasars
preventing a reliable estimate of the relative contributions of O~I
and Si~II to the blend even in studies of high quality spectra
dedicated to Si~II (e.g. Appendix A in Baldwin et al. 1996). In \zw\
the Si~II~$\lambda 1309$ component of the Si~II doublet is clearly
resolved.  The Si~II~$\lambda 1304$ component is hopelessly blended
with the O~I~$\lambda 1304$ line, which is actually a triplet at
1302.17, 1304.86, and 1306.0~\AA. However, the Si~II 1309/1304 flux
ratio should either be 2/1 (optically thin case), or 1/1 (optically
thick case), and thus the clear detection of the 1309 component allows
an estimate of the 1304 component flux within a factor of two.  The
template fit to the O~I+Si~II blend (Fig.3), which assumes an
optically thick Si~II doublet, suggests that 0.5 of the blend flux is
due to O~I. If the Si~II doublet is optically thin (less likely given
the results in \S\S3.1.5, 3.1.6), then the Si~II~$\lambda 1304$
component is only half as strong, and then 0.56 of the blend flux is
due to O~I. Using the O~I~$\lambda 8446$ flux from Rudy et al. then
gives $R_{\rm O~I}\equiv \lambda 8446/\lambda 1304$ photon flux ratio
is 5.5 (4.8), for the optically thick (thin) case.

The shortest wavelength flux in the spectrum from which the Rudy et
al. O~I~$\lambda 8446$ flux is taken (at 7500\AA) agrees within $\sim
20$\% of the flux in the optical spectrum shown in Fig.1, thus
variability between the times the $\lambda 8446$ and the $\lambda
1304$ lines were measured is most likely not a significant source of
error.

Kwan \& Krolik (1981) have found that $R_{\rm O~I}$ can increase from
unity to 1.3 due to Balmer continuum absorption of $\lambda 1304$
photons, and the production of $\lambda 8446$ photons by collisional
excitation of O~I ions in the $3s\,^3\!S$ level. A third mechanism
which can increase $R_{\rm O~I}$, noted by Grandi (1983), is the
significant probability for a $2p\!^4\,^1\!D-3s\,^3\!S$ $\lambda 1641$
semi-forbidden transition for $\tau_{\lambda 1304}\ge 10^5$, or
equivalently $\tau_{Ly\alpha}\ge 10^9$, which would further increase
$R_{\rm O~I}$. Kwan (1984) included the O~I]~$\lambda 1641$ transition
and indeed found $R_{\rm O~I}\sim 3.2$, where most of the increase
from unity is due to the ``re-routing'' of $\lambda 1304$ photons to
$\lambda 1641$.

This mechanism can be tested in \zw\ since the He~II~$\lambda 1640$
line is strongly blueshifted (to $\sim 1630$~\AA), while the O~I lines
are not.  This allows us to look for the O~I]~$\lambda 1641$ line and
test if it can indeed explain the very high $R_{\rm O~I}$ observed. No
strong line is detected at 1641~\AA, and we estimate that such a line
cannot add more than $\sim 30$\% to the observed $\lambda 1304$ line.
The absence of a significant flux in the O~I]~$\lambda 1641$ line
therefore implies that most of the
\lyb\ absorption by O~I in \zw\ takes place at a depth where
$\tau_{Ly\alpha}< 10^9$.

Rudy et al. gave observational evidence that collisional excitations
cannot significantly increase $R_{\rm O~I}$ in \zw. Thus, of the three
mechanisms mentioned above, the only one left is Balmer continuum
absorption of $\lambda 1304$ photons. It remains to be explored
whether this mechanism can suppress the $\lambda 1304$ line by the
required factor of $\sim 5$ when $\tau_{Ly\alpha}< 10^9$.

Another option is that there are few O~I~$\lambda 1304$ photons because the 
observed emission is strongly reddened,
as discussed below in \S 3.3.

Another observed emission line from O is the O~IV]$~\lambda 1402$
line. This line is a multiplet of five components and it is strongly
blended with the Si~IV~$\lambda 1397$ doublet producing a single broad
$\lambda 1400$ feature in most AGNs. The relative contribution of O~IV
and Si~IV has been debated in the past (Baldwin \& Netzer 1978; Wills
\& Netzer 1979; Tytler \& Fan 1992; Wills et al. 1993; Laor et
al. 1994a). The narrow lines of \zw\ appear to provide a good
opportunity to determine accurately the relative contribution of O~IV
and Si~IV to the $\lambda 1400$ feature.

The peak of the $\lambda 1400$ blend is clearly split in \zw\ into two
peaks at 1392.1~\AA\ and 1400.6~\AA. These correspond well to the
expected positions of the Si~IV doublet at 1393.76~\AA\ and
1402.77~\AA, blueshifted by $\sim 2$~\AA.  However, some O~IV] must
also be present based on the following argument.  The 1392.1~\AA\
feature must be due to Si~IV only (the shortest wavelength O~IV]
component is at 1397.2~\AA). The second Si~IV doublet component is at
1400.6~\AA\ and its flux should either be equal to (optically thick)
or half of (optically thin) the 1397.2~\AA\ line flux. The observed
1400.6~\AA\ component is stronger than the 1392.1~\AA\ component, and
must therefore include some O~IV] emission.  A specific template
deblending is shown in Fig.3, using \ha\ as a template, which implies
a Si~IV/O~IV] flux ratio of 1/2.  However, this ratio is not well
constrained since the intrinsic line profile is probably not
symmetric, as suggested by the large observed blue excess flux, and
the fact that all other high ionization lines display a strong blue
excess flux (see \S 3.2). Thus, we can only verify that both Si~IV and
O~IV] are present, and that they have roughly comparable flux.

\subsubsection{Mg}

The Mg~II ion is produced by photons above 7.646~eV and destroyed by
photons above 15.035~eV, it therefore survives mostly in the partially
ionized H region, where photons above 13.6~eV are blocked. The
Mg~II~$\lambda 2798$ doublet is therefore a diagnostic of the
conditions in the extended partially ionized layers inside the BLR
clouds. The flux ratio of the Mg~II 2796.35~\AA/2803.53~\AA\
components indicates the optical depth in the line. Below we describe
the observed flux ratio and discuss its implication.

The Mg~II~$\lambda 2798$ profile is clearly double peaked. Its core is
well fit with a 1.2/1 ratio for the 2796.35~\AA/2803.53~\AA\ doublet
lines using the \ha\ profile as a template (Fig.3). A good fit
requires that the actual continuum level be placed 20\% below the
observed flux level at the apparent line base. This continuum level
matches very well the continuum level inferred at 2800~\AA\ from a
linear interpolation of the continuum at 2650~\AA\ and 3020~\AA.

What does the 1.2/1 doublet ratio implies for the conditions in the
Mg~II emitting gas?  The 2803.53~\AA, 2796.35~\AA\ lines correspond to
transitions from the $3p\,^2\!P_{1/2}$ and $3p\,^2\!P_{3/2}$ levels to
the ground $3s\,^2\!S_{1/2}$ level. The collision strength at
$T=10^4$~K is $\Omega(^2P,^2S)=16.9$ (Pradhan \& Peng 1995), yielding
collisional de-excitation rates of
$q(3p\,^2\!P_{3/2},3s\,^2\!S_{1/2})=3.65\times 10^{-7}$ and
$q(3p\,^2\!P_{1/2},3s\,^2\!S_{1/2})=7.3\times 10^{-7}$.  The radiative
de-excitation rates for the $3p\,^2\!P_{3/2,1/2}$ levels are
$A=2.6\times 10^8$~s$^{-1}$ (Morton 1991). Collisional de-excitation
dominates when
\[\tau_0 n_e>A/q,\] 
where $\tau_0$ is the line center optical depth for resonance
scattering, i.e when $\tau_0 n_e>7.1\times 10^{14}$~cm$^{-3}$ and
$3.6\times 10^{14}$~cm$^{-3}$ for the $3p\,^2\!P_{3/2}$ and
$3p\,^2\!P_{1/2}$ levels. For $\tau_0 n_e>7.1\times 10^{14}$~cm$^{-3}$
the Mg~II 2796.35~\AA/2803.53~\AA\ doublet ratio is thermalized,
i.e. it changes from 2/1 to 1/1 (see Hamann et al. 1995 for results on
other analogous UV doublets). Thus, the observed 1.2/1 ratio implies
that most of the Mg~II emission is produced in a region where $\tau_0
n_e\gtrsim 10^{15}$~cm$^{-3}$.

What does the above result implies for the conditions in the partially
ionized BLR gas?  As shown below, the value of $\tau_0 n_e$ provides a
direct measure of the distance of the BLR clouds from the central
ionizing continuum source. The line center optical depth is
$\tau_0=\sigma_{\nu_0}\Sigma_{\rm Mg~II}$ where
$\sigma_{\nu_0}=1.47\times 10^{-12}$~cm$^2$ is the total Mg~II doublet
line center absorption cross section (for $T=10^4$~K), and
$\Sigma_{\rm Mg~II}$ is the column density of Mg~II ions.  The
partially ionized H region, where Mg~II survives, has a column density
about 10-30 times larger than the column density of the highly ionized
H region, given by $\Sigma_{\rm H~II}=10^{23}U$~\cmII, where $U$ is
the ionization parameter ($\equiv n_{\gamma}/n_e$, the photon/electron
density).  Balmer continuum absorption may limit the depth of the
Mg~II doublet emitting layer, but this absorption will not affect our
conclusion.  Using the Mg/H solar abundance ratio of $3.8\times
10^{-5}$, and assuming most Mg is Mg~II, we get $\Sigma_{\rm
Mg~II}\simeq 4\times 10^{19}U$~\cmII, or $\tau_0=6\times 10^7U$,
assuming the gas is ionization, rather than matter bounded. The Mg~II
doublet is therefore thermalized when $6\times 10^7Un_e\gtrsim
10^{15}$, or $n_{\gamma}\gtrsim 10^7$~\cmIII.

Reverberation mapping (e.g. Netzer 1990; Peterson 1993; Maoz 1995;
Kaspi et al. 1996), and theoretical considerations (Netzer \& Laor
1993) suggest $R_{\rm BLR}=0.1L_{46}^{1/2}$~pc, which implies
$n_{\gamma}\sim 2\times 10^9$~\cmIII in the BLR.  Thus, the observed
1.2/1 Mg~II doublet ratio provides direct confirmation that the BLR is
indeed optically thick in the Mg~II~$\lambda 2798$ line, as predicted
by `standard' photoionization models for the BLR.  More refined
calculations using actual photoionization models can be used to
explore the exact doublet ratio predicted by various BLR models and
its agreement with the observed value.

Note that one generally expects a doublet ratio from optically thick
gas to be somewhere between 2/1 and 1/1, rather than 1/1, since (as is
always the case in atmospheres), most of the contribution to the line
flux arises from the region where the emission just becomes optically
thick, i.e.  from the region where the doublet ratio changes from 2/1
to 1/1.

\subsubsection{Al}

The components of the Al~III~$\lambda 1857$ doublet
are widely spaced ($\sim 1300$\kms) and are therefore clearly
resolved in \zw. Al~III is created by photons above 18.829~eV, and 
destroyed by photons above 28.448~eV, thus unlike Mg~II, it exists in 
the highly ionized H region only, and the observed Al~III doublet ratio
provides constraints on the BLR H~II region, as discussed below.

The template fit suggests a 1.25/1 ratio for the
1854.71~\AA/1862.79~\AA\ doublet lines (Fig.3).  This ratio is
somewhat uncertain since there are significant deviations from the
\ha\ template fit, possibly due to blending with Fe~II multiplets
(e.g. Wills, Netzer \& Wills 1980), or an intrinsic blue excess flux,
as displayed by other high ionization lines (\S 3.2). Baldwin et
al. (1996) found a similar 1/1 flux ratio in three objects where the
Al~III doublet was clearly detected.

Below we calculate the expected Al~III~$\lambda 1857$ 
doublet ratio, as calculated above for the Mg~II $\lambda 2798$ doublet.

The Al~III and Mg~II ions are both Na~I like.
Thus, the 1854.71~\AA\ and 1862.79~\AA\ Al~III lines are also produced 
by transitions from the $3p\,^2\!P_{1/2}$ and $3p\,^2\!P_{3/2}$
levels to the ground $3s\,^2\!S_{1/2}$ level. The collision strength
at $T=10^4$~K is $\Omega(^2P,^2S)=16.0$ (Dufton \& Kingston 1987), 
the radiative de-excitation rate is $A=5.4\times 10^8$~s$^{-1}$ 
(Morton 1991), the solar Al/H abundance ratio is $3.6\times 10^{-6}$,
and $\sigma_{\nu_0}=9.4\times 10^{-13}$~cm$^2$.
Al~III exists in the highly ionized H region only, 
where $\Sigma_{\rm H~II}=10^{23}U$~\cmII. Using these parameters,
and assuming Al is mostly Al~III, 
we find that the Al~III~$\lambda 1857$ doublet is thermalized
for $n_{\gamma}\gtrsim 4\times 10^9$~\cmIII. 

The observed 1.25/1  Al~III doublet ratio suggests that
Al~III is formed in clouds at, or inside the `standard'
BLR radius of $0.1L_{46}^{1/2}$~pc, where 
$n_{\gamma}\gtrsim 2\times 10^9$~\cmIII. The fact that the Al~III doublet 
is expected to be only marginally thermalized at the `standard' BLR radius
suggests that this doublet ratio can provide an interesting probe of the
size of the BLR.

The relatively large Al~III~$\lambda 1857$ EW (4.4~\AA) requires a
BLR component where Al~III is the dominant ionization state of Al,
which occurs for $U\sim 10^{-2}-10^{-3}$ (e.g. Baldwin et al.
1996, Fig.7a). 
Since $n_{\gamma}\gtrsim 4\times 10^9$~\cmIII, the above range
in $U$ implies 
$n_e\gtrsim 4\times 10^{11}-4\times 10^{12}$~\cmIII, which is comparable,
or denser, than the component responsible for the C~III]~$\lambda 1909$ 
emission (\S 3.1.2).

We caution that resonance scattering of continuum photons has not
been included in the above discussion. This process could contribute an 
EW of $\sim 2$~\AA\ to each of the Al~III~$\lambda 1857$ doublet components, 
i.e. as much as observed, if the velocity dispersion in the 
BLR clouds is as large as the lines FWHM, and the clouds cover
$\gtrsim 30$\% of the continuum source (e.g. Ferguson, Ferland 
\& Pradhan 1995). 
Continuum resonance scattering would also imply a much lower
value for $\tau_0$ than estimated above (due to the large velocity 
gradients required in the BLR clouds). Calculations with detailed
photoionization codes are required in order to understand 
whether continuum resonance scattering is indeed viable.

\subsubsection{Si}

The Si~II 1260.42~\AA, 1264.74~\AA, 1265.00~\AA, 1304.37~\AA, 
1309.28~\AA, 1808.01~\AA, 1816.93~\AA, and 1817.45~\AA\ lines are
clearly detected. The flux of the Si~II $\lambda 1194$ and $\lambda 2335$ 
blends is well constrained. This wealth of data provides a 
unique opportunity to explore the Si~II line formation mechanisms
(e.g. Baldwin et al. 1996, Appendix C). Such analysis is beyond the 
scope of this paper. 
 
The outer shell electronic configuration of the Si~III ion is
analogous to that of C~III. The analogous line to the [C~III]~$\lambda
1907$ line (\S 3.1.2) is the [Si~III]~$\lambda 1883$ line, which can
also be used as a density diagnostic for the Si~III gas. Weak [Si
III]~$\lambda 1883$ emission may be present at 1881.1~\AA.  The
identification of this marginal feature is supported by the fact that
its 250~km~s$^{-1}$ blue shift matches the Si III]~$\lambda 1892$
shift, and by the possible presence of the analogous [C III]~$\lambda
1907$ line. As for the C~III ion, the $R_{\rm Si~III}$
$\equiv$~[Si~III]~$\lambda 1883/$ Si~III]~$\lambda 1892$ flux
ratio is a robust density indicator (Keenan, Feibelman \& Berrington
1992). The observed $R_{\rm Si~III}\lesssim 0.1$ implies that some of
the Si~III resides in gas with $n_e\sim 5\times 10^5$~\cmIII, i.e. the
same conditions implied by $R_{\rm C~III}$, as expected since the
spatial distribution of both ions in photoionized gas should largely
overlap.

The Si~III]~$\lambda 1892$/C~III]~$\lambda 1909$ flux ratio of $\sim 3.5$
is significantly larger than the typical ratio
of $\sim 0.3\pm 0.1$ observed in quasars (Laor et al. 1995). This
high ratio results from the factor of 5-10 suppression in the 
C~III]~$\lambda 1909$ flux (\S 3.1.2), rather than significant
enhancement of Si~III]~$\lambda 1892$. The most likely interpretation 
for this high ratio is a relatively dense BLR.
The critical density for C~III]~$\lambda 1909$,  
$n_c\sim 5\times 10^9$~\cmIII, is significantly smaller than 
for Si~III]~$\lambda 1892$, where $n_c=1.1\times 10^{11}$~\cmIII.
Thus, if the BLR density is $10^{11}$~\cmIII, rather than $10^{10}$~\cmIII,
C~III]~$\lambda 1909$ would be suppressed by a factor of $\sim 10$,
while Si~III]~$\lambda 1892$ would not be affected, consistent with
our observations. 

\subsubsection{Fe}

The positions of some Fe~II UV blends are marked in Fig.1 
only to illustrate various possible Fe~II emission features.
Numerous additional weak Fe~II blends are 
most likely present at other wavelengths.
Due to the extreme complexity of the Fe~II ion we do
not attempt to identify individual features. 

The likely presence of Fe~II emission well below 2000~\AA, and
possibly even at $\sim 1110$-1130~\AA, indicates that Fe~II is
excited by processes other than just collisions, as the electron
temperature in the Fe~II region is most likely 
far too low for significant
collisional excitation of levels $\gtrsim 10$~eV above the ground state. 
One such process,
suggested by Penston (1987), is resonance absorption of \lya\
photons by Fe~II. Johansson \& Jordan (1984) found \lya\ 
resonance absorption to be significant in various stellar systems,
and identified the Fe~II~$\lambda 1294$,  
Fe~II~UV~191~$\lambda 1787$, and the Fe~II~$\lambda 1871$ multiplets
as the signatures of such a process (see also Johansson \& Hamann 1993). 
The Fe~II~UV~191 multiplet is very
prominent in \zw, and weak emission blends are clearly apparent
at 1294~\AA, and 1870~\AA, suggesting that resonance absorption
of \lya\ photons may be a significant excitation mechanism
for Fe~II in \zw\ as well. Resonance scattering of continuum photons, 
and Fe~II-Fe~II line flouresence could also be 
a significant process for populating high lying Fe~II levels
(e.g. Netzer \& Wills 1983; Wills, Netzer \& Wills 1985).
Clearly, the Fe~II rich spectrum of \zw\ should serve as a 
valuable tool for future studies of Fe~II emission in AGNs.

A number of Fe~III multiplets are clearly present.
The Fe~III~UV~34 multiplet, clearly identified here, was
first discovered by Hartig \& Baldwin (1986) in a 
broad absorption line quasar. An emission feature near $\sim 2070$~\AA\
is commonly seen in quasars (e.g. Wills et al. 1980), 
and was identified as a likely Fe~II blend. Baldwin et al. (1996)
suggested that this feature is due to Fe~III~UV~48 emission. Here the
three Fe~III~UV~48 components at 2062.2~\AA, 2068.9~\AA, and 2079.65~\AA\
are clearly resolved, which verifies the identity of this feature.
The Fe~III~UV~47~$\lambda\lambda 2419.3, 2438.9$ doublet is most likely 
present as well.
Other Fe~III blends, such as Fe~III~UV~50 and Fe~III~UV~68,
and in particular the resonance Fe~III~UV~1 at 1122-1130~\AA\ 
may also be present, but these identifications cannot be verified
here since these blends are not clearly resolved.

The relative flux ratio of the multiplet components can provide
an important diagnostic for the optical depth and excitation mechanism 
of Fe~III. Given the complexity of the Fe~III ion, these diagnostics
are beyond the scope of this paper.

\subsection{Line Profiles and Velocity Shifts}

Optical spectra of \zw\ revealed two velocity systems. 
The first system is at $z=0.0608$ for the low ionization forbidden lines 
and Balmer lines,
including \ha, \hb, optical Fe~II multiplets, [Ca~II] $\lambda\lambda 7291.46,
7323.89$,~~~~[S~II]~~~$\lambda\lambda 4068.60,4076.35$, [N~II]~$\lambda
5754.57$,  He~I $\lambda 5875.6$, Na~I $\lambda\lambda 5889.95,$ 5895.92,
and [O~I] $\lambda\lambda 6300.3, 6363.8$. The second 
system is at $z=0.0587$ (i.e. blueshifted by $\sim 630$~\kms) for the 
higher ionization forbidden lines,
[O~III] $\lambda\lambda 4958.82,5006.85$ and 
[Ne~III] $\lambda\lambda 3868.74,3967.47$ 
(Phillips 1976, Oke \& Lauer 1979).

We find the same trend in the UV. As shown in Table 1, the low 
ionization permitted lines, including O~I $\lambda 1304$, 
C~II $\lambda 1335.3$ and 
C~II] $\lambda\lambda 2324.21, 2328.84$, 
N~II] $\lambda\lambda 2139.68, 2143.45$,
Al~II $\lambda 1670.89$ and 
Al~II] $\lambda 2669.95$, the various Si~II multiplets,
and Fe~II~191~$\lambda 1787$ are all blueshifted by 
$\lesssim 200$\kms\ with respect to $z=0.0608$. 
Note that shifts of $\lesssim 100$\kms\ are consistent with
zero shift, given the accuracy of our wavelength 
zero point calibration using the 
interstellar medium lines. Higher ionization 
lines, including N~III] $\lambda 1750$, 
Al~III $\lambda\lambda 1854.72, 1862.79$,
Si~III] $\lambda 1892.03$ and Si~IV $\lambda\lambda 1393.76, 1402.77$
are blueshifted by $\sim 300-500$\kms, while the
highest ionization lines C~IV~$\lambda\lambda 1548.19, 1550.77$ and
N~V $\lambda\lambda 1238.8, 1242.8$, are blueshifted by $\sim 900$\kms.
The highest blueshift, $\sim 2000$\kms, 
is displayed by He~II~$\lambda 1640.4$.

This trend of increasing blueshift with increasing ionization level is 
observed in most quasars 
(e.g. Gaskell 1982; Wilkes 1984, 1986; Uomoto 1984; 
Espey et al. 1989; Corbin 1990). However, the typical amplitude of the
high ionization line blueshift, such as C~IV and N~V,
 is only $\sim 200-300$\kms,
and for He~II it is $\sim 500$\kms\ (Tytler \& Fan 1992; Laor et al. 1995), i.e.
about four times lower than found in \zw.

The low ionization line profiles are mostly consistent with the rather
symmetric \ha\ profile.  With increasing blueshift the lines get
progressively broader and develop a progressively stronger blue excess
asymmetry (see Fig.3).
  
\subsection{Reddening}

As shown in Fig.1, the optical-UV spectrum of \zw\ is significantly redder 
than observed in typical quasars. Could the intrinsic continuum of \zw\ be 
similar to the Francis et al. composite, and the observed red continuum just 
be due to reddening? 

Intrinsic reddening by Galactic like dust is ruled out as it would
have produced a very pronounced broad absorption feature at
1800-2500~\AA, which is not observed.  The grain size distribution in
AGNs may, however, be different than in the Galactic ISM (e.g. Laor \&
Draine 1993), and in particular it may not produce a significant
2200~\AA\ extinction bump. Thus, we cannot securely rule out reddening
just based on the absence of the 2200~\AA\ feature.

Reddening would also affect various optical/UV line ratios. In
particular, assuming the correct optical/UV flux distribution is given
by the Francis et al.  composite implies that $R_{\rm O~I}\sim 1.2$
(instead of 5.5), which is consistent with the value predicted by the
Kwan \& Krolik (1981) calculation (which does not include the
production of a O~I]~$\lambda 1641$ shown here to be insignificant).
Dereddening also affects the \lya/\hb\ flux ratio. The intrinsic ratio
would then be 32 instead of 7.6. Such a high \lya/\hb\ ratio is
theoretically possible for plausible BLR parameters (e.g. Netzer et
al. 1995), but it is much higher than the ratio of 5-15 typically
observed in quasars.

If the intrinsic continuum of \zw\ is similar to the Francis et al.
composite, and the observed continuum is due to reddening by Galactic
like dust (except the 2200~\AA\ bump) intrinsic to \zw, then the
implied reddening is of E(B-V)$\sim 0.2$. For a Galactic dust/gas
ratio this implies $N_{\rm H}=10^{21}$~\cmII. The ROSAT PSPC spectrum
of \zw\ implies an intrinsic neutral column density of no more than
$(1.5\pm 0.7)\times 10^{20}$~\cmII\ (see \S 4.2). Thus, the gas
associated with the dust must either be highly ionized ($U>1$) to
prevent strong X-ray absorption below 0.4~keV, or the dust/gas ratio
needs to be at least 5 times larger than the Galactic value. The X-ray
absorption by the grain component itself is negligible at the above
columns (see Fig.6 in Laor \& Draine 1993) and therefore cannot be
used to constrain the dust column.

A nearly featureless X-ray power-law may still be associated with a
large dust column density along the line of sight if most of the
observed X-rays are electron scattered by an extended medium. In this
case only a small fraction of the X-rays would be scattered and thus
although the X-ray power-law would be featureless, its normalization
relative to the optical would be significantly reduced. However,
Lawrence et al. (1997) find an optical to X-ray power-law of
$\alpha_{ox}=-1.41$, which is very close to the average value of
$\alpha_{ox}=-1.48$ for radio quiet quasars (Laor et al. 1997). Thus,
it appears most likely that the primary X-ray source in \zw\ is
observed directly, unless the direct optical flux is also completely
obscured.

Theoretical~ and ~observational~uncertainties   \hbox{therefore} do not allow us
to clearly conclude whether \zw\ is reddened, or not.

\section{Intrinsic Absorption}
In this section we describe observational evidence for associated UV 
absorption and 
the implied constraints on the absorber column. We also discuss constraints 
based on the soft X-ray spectrum, and a possible relation of the absorber
to a highly blueshifted emission component.

\subsection{UV absorption}

\subsubsection{Observational evidence}

Figure 4 displays evidence for associated absorption in \lya, N~V, and C~IV. 
The clearest evidence for absorption is seen in the 
N~V~$\lambda\lambda 1238.8,$ 1242.8 doublet. The N~V absorption doublet is
blueshifted by $\sim 1870$\kms\ relative to the $z=0.0608$ frame, it has
a FWHM$\sim 300$\kms, and an absorption EW of $\sim 0.25$~\AA\ (determined
using the $\lambda 1242.8$ component). Since the spectral resolution of the
FOS at N~V is $\sim 230$\kms, the absorption system may not be truly 
resolved. The C~IV~$\lambda\lambda 1548.19, 1550.77$ profile is consistent
with having a very similar absorption system, although the spectrum has a very 
low S/N at the blue wing of C~IV (falls on the blue edge of the G190H).
The \lya\ profile appears to be affected by an absorption system with 
the same velocity shift, and about the same EW and FWHM.

The presence of high ionization lines, and lack of low ionization lines
suggests the absorber is associated with \zw, rather
than with an intervening system unrelated to \zw\ (see e.g. Hamann 1997).
Most of the constraints below are independent of the exact location of the
absorber.

\subsubsection{Implied constraints}

The column density associated with the observed absorption EW is a
function of the absorber optical depth. The optical depth can be
deduced from the ratio of EW of the two doublet components, even when
the absorption line profiles are not resolved, assuming the absorber
completely covers the emission source. In the optically thin limit
this ratio equals the ratio of oscillator strengths, i.e
EW(1238.8)/EW(1242.8)$=2$, and when both absorption lines are
saturated this ratio approaches unity. The observed ratio is $\sim
1.4$, but it is rather uncertain since the 1242.8~\AA\ component
absorbs the line peak where the underlying emission is not clearly
defined, and the 1238.8~\AA\ component is partly blended with Galactic
Si~II~$\lambda 1304.37$ absorption. The absorption optical depth
therefore remains uncertain.

In the optically thin limit the absorber column density is related to the EW
through
\[ N_{\rm abs}=1.13 \times 10^{20}\lambda^{-2} f^{-1}{\rm EW}~{\rm cm}^{-2},\]
where $\lambda$, the absorption wavelength, and its EW are in units of 
~\AA, and $f$ is the oscillator strength. 
Using $f(1215.67)=0.416, f(1550.77)=0.095$,  
$f(1242.8)=0.078$ and an absorption EW of 0.25~\AA\ for all lines, 
we get $N_{\rm H~I}=4.6\times 10^{13}$~cm$^{-2}$,
$N_{\rm C~IV}=1.2\times 10^{14}$~cm$^{-2}$, and
$N_{\rm N~V}=2.3\times 10^{14}$~cm$^{-2}$. These values are lower limits on
the absorber columns. 

The upper limits on the columns are obtained in the optically thick
limit, i.e. when only thermal broadening is present.  If the
absorption line profile is not resolved, and only the absorption EW is
given, then the deduced column density depends on whether the line is
optically thick in the Doppler core only, or whether it is also thick
in the Lorentzian wings (i.e. ``damped absorption''). The first case
applies when
\[ {\rm EW}/\Delta\lambda_D\lesssim 6, \]
where 
\[ \Delta\lambda_D=4.285\times 10^{-5} \lambda T_4^{1/2}A^{-1/2} \]
is the thermal line width in \AA, 
$T=10^4T_4$~K, and $A$ is the atomic weight.
The absorber column density is then given by
\[ N_{\rm abs}=
2 \times 10^{20}\lambda^{-2} 
f^{-1}\Delta\lambda_D e^{({{\rm EW}/2\Delta\lambda_D})^2}. \]
When absorption is dominated
by the Lorentzian wings, i.e. when 
\[ EW/\Delta\lambda_D\gtrsim 6, \]
The absorber column density is
\[ N_{\rm abs}\simeq 
3.35 \times 10^{39}\lambda^{-4}f^{-1}\Gamma^{-1}{\rm EW}^2, \]
where $\Gamma$ is the radiative decay rate 
($\Gamma_{1215.67}=6.265\times 10^8~s^{-1},
\Gamma_{1550.77}=2.64\times 10^8~s^{-1},
\Gamma_{1242.80}=3.387\times 10^8~s^{-1}$). 
 Assuming $T_4=2$, the typical
temperature of photoionized gas where C~IV or N~V dominate, we get the 
maximum possible columns of
$N_{\rm H~I}=4.3\times 10^{14}$~cm$^{-2}$ 
(where ${\rm EW}/\Delta\lambda_D< 6$)
$N_{\rm C~IV}=1.4\times 10^{18}$~cm$^{-2}$, and
$N_{\rm N~V}=3.3\times 10^{18}$~cm$^{-2}$ 
(both cases ${\rm EW}/\Delta\lambda_D> 6$).
The H~I column density upper limit is rather uncertain since a Doppler core 
absorption gives 
$N_{\rm abs}\propto e^{ ({\rm EW}/2\Delta\lambda_D)^2 }$, and thus 
a likely $\pm 30$\% error in the H~I EW corresponds to an
upper limit in the range of $0.1-3.1\times 10^{15}$~cm$^{-2}$. The
C~IV and N~V upper limits are uncertain by only a factor of $\lesssim 2$, 
since $N_{\rm abs}\propto {\rm EW}^2$, when the Lorentzian wings dominate 
the absorption.

To infer the total H column density from the H~I column density 
one needs to know
the H ionization state. The 
H~II/H~I fraction in the absorbing gas is related to the ionization
parameter through 
\[ N_{\rm H~II}/N_{\rm H~I}\simeq 2\times 10^5 U \] 
(e.g. Netzer 1990). Thus, the upper limit on the H column density is
\[ N_{\rm H}\simeq N_{\rm H~II}=(0.2-6)\times 10^{20}U~{\rm cm}^{-2}.\]
The presence of N~V and C~IV ions suggests that the absorber has
$U\sim 0.01-1$, thus the H column density upper limit is probably in
the range $10^{17}\lesssim N_{\rm H} \lesssim 6\times 10^{20}$.  If
the absorber metal abundance is solar then the above $N_{\rm H}$
implies metal columns upper limits of $3.6\times 10^{13}\lesssim
N_{\rm C} \lesssim 2\times 10^{17}$ and $10^{13}\lesssim N_{\rm N}
\lesssim 6.7\times 10^{16}$. These values overlap with the directly
determined constraints on the C~IV and N~V columns and suggests that
significant fractions of C and N are indeed in the form of C~IV and
N~V, consistent with the assumption of $U\sim 0.01-1$. Detection of
O~VI~$\lambda\lambda 1031.93, 1037.63$ absorption would allow a much
tighter constraint on the absorber's $U$.

\subsection{Soft X-ray absorption}

In addition to the UV absorption lines there may also be some
soft X-ray absorption in \zw. Boller, Brandt, \& Fink (1996) 
and Lawrence et al. (1997) fit a 
single power-law to the {\em ROSAT} PSPC spectrum of \zw\ and find
a best fit column density of 
$N_{\rm H~I}=(6.5-6.7\pm 0.7)\times 10^{20}$~cm$^{-2}$ (90\% error) versus
a Galactic value of $(5.0\pm 0.1)\times 10^{20}$~\cmII\ (Elvis et al.
1989), suggesting that there may be an intrinsic 
absorber in \zw\ with   
$N_{\rm H}\sim (1.5-1.7\pm 0.7)\times 10^{20}$~cm$^{-2}$. 
Lawrence et al. (1997) find that a fit including only Galactic absorption 
can be excluded with a confidence level greater than 99\%.

The observed total column density in \zw\ is optically thick below
0.38~keV, where He dominates the opacity. Thus, the only direct
constraint from {\em ROSAT} is for $N_{\rm He~I}\sim (1.5\pm
0.7)\times 10^{19}$~cm$^{-2}$, assuming the cosmic $n_{\rm He}/n_{\rm
H}=0.1$ abundance.  A $\sim 20$\% higher column density applies if
most of He is He~II, as deduced using the Verner \& Yakovlev (1995)
He~I and He~II absorption cross sections.

Associations of UV and X-ray absorbers with the same gas was claimed
in 3C~351 (Mathur et al. 1994), and NGC~5548 (Mathur, Elvis \& Wilkes
1995), but disputed in NGC~3516 (Kriss et al. 1996a, 1996b).  Can the
UV and the X-ray absorption in \zw\ originate in the same absorber?
In a cloud photoionized by a typical AGN continuum there is a surface
layer of thickness $N_{\rm H}\sim 10^{23}U$~cm$^{-2}$ where H is
mostly ionized. At the surface of this ``H~II'' layer there is a
thinner ``He~III'' layer.  For a typical AGN continuum the column
density of the ``He~III'' layer is $N_{\rm H}({\rm He~III})\sim
2\times 10^{21}U$~cm$^{-2}$. Thus, irrespective of the value of $U$,
the ``He~III'' layer is about three times thicker than the upper limit
of $N_{\rm H}\le 6\times 10^{20}U$~cm$^{-2}$ deduced from the UV
absorber in \zw.  The He in the UV absorber would be mostly He~III,
and the soft X-ray absorption would originate in the residual He~II
within the UV absorber. The He~III/He~II fraction within the
``He~III'' region is given by $N_{\rm He~III}/N_{\rm He~II}\simeq 10^3
U$ (Netzer 1990).  The upper limit on the He~II column density in the
UV absorber is then
\[ N_{\rm He~II}=N_{\rm H}\times\frac{N_{\rm He}}{N_{\rm H}}
\times\frac{N_{\rm He~II}}{N_{\rm He}} 
\le 6\times 10^{20}U\times 0.1\times (10^3U)^{-1},\]
i.e. $N_{\rm He~II}=6\times 10^{16}~{\rm cm}^{-2}$
which is $\sim 300$ smaller than the $N_{\rm He~II}$ required to produce
the X-ray absorption. It thus appear that the upper limit on the column
of the layer which produces the UV absorption is much smaller than required
to produce the possible X-ray absorption.

This line of argument suggests that the UV and X-ray absorbers are
distinct. However, given the uncertainty in the upper limit on the
\lya\ absorption EW, which sets $N_{\rm H}$, and the uncertainty in
the far UV spectral shape, which sets $N_{\rm He~III}(U)$, we cannot
conclusively rule out the possibility that an ``He~II'' region is
present behind the ``He~III'' region.  Such an ``He~II'' region can
produce a high He~II column, which would be required if the X-ray
absorption is real, together with a relatively small total H~I column,
implied by the UV absorption lines.

\subsection{UV and optical emission}

The UV absorber blueshift of $\sim 1870$\kms\ is remarkably close to
the He~II~$\lambda 1640$ emission line blueshift of $\sim 1990$\kms,
suggesting that the strongly blueshifted He~II emission line peak may
originate in the same outflowing gas which produces the UV absorption
lines. In order to emit the observed He~II~$\lambda 1640$ flux the
absorber must absorb most photons just above the He~II bound-free edge
at 4~Rydberg. As described above, when $N_{\rm H}\sim 6\times
10^{20}U$~cm$^{-2}$ the UV absorber is within a factor of three of the
column density required to form an ``He~II'' region, i.e. it absorbs
$\sim 1/3$ of He~II ionizing photons, and thus if the column density
of the UV absorber is close to the upper limit, and its covering
factor is close to unity, it may contribute significantly to the
observed He~II~$\lambda 1640$ emission.

The intrinsic X-ray absorber becomes optically thick below 0.2~keV.
It is therefore highly optically thick at 4~Rydberg and can
produce the observed He~II~$\lambda 1640$ emission if its covering factor
is reasonably large ($>10$\%).

It is interesting to note that Oke \& Lauer (1979) found evidence for 
[O~III]~$\lambda 5007$ and [Ne~III] $\lambda 3868$ emission components at
$z=0.0548$ which corresponds to a $\sim 1800$\kms\ blueshift with
respect to the $z=0.0608$ used here. The same feature was also noted by
van Groningen (1993) in a high spectral resolution profile of
[O~III] $\lambda 5007$. If these emission components also originate 
in the UV absorber then their flux can be used as an additional constraint
on the density, column density, and ionization parameter of the UV absorber.

\subsection{Future observations}

Higher spectral resolution observations of the \lya\ N~V and C~IV
absorption lines would allow a much more accurate determination of the
absorption profile, and thus of the absorber's column. If the H~I
column density is indeed close to the upper limit, as required to
``unite'' the UV absorption, X-ray absorption, and He~II emission,
then the \lya\ absorption profile, for example, would appear as a
$\sim 60$\kms\ wide, steep sided, and flat bottomed trough going down
to zero flux. The metal absorption lines should also be resolved as
deep narrow optically thick absorption lines with damping wings. If
the absorption lines remain shallow and broad then the lower limits to
the absorbing columns would apply. The metal lines absorption doublet
ratios can be used to establish the covering factor of the UV
absorber.

Higher S/N X-ray spectra below 0.5~keV are required to establish whether
the soft X-ray absorption hinted at by the {\em ROSAT} PSPC is indeed real.

\section{Discussion} 

The strong optical Fe~II emission of \zw, its weak [O~III] emission, 
strong IR emission, and ``red'' UV continuum are all typical properties of
low ionization broad absorption line quasars (BALQSOs; Weymann et al. 1991, 
Boroson \& Meyers 1992; Sprayberry \& Foltz 1992). The presence of weak UV 
absorption at a blueshift of $\sim 2000$\kms\ 
in \zw\ suggests it may be a ``failed'' low ionization BALQSO, i.e.
a BALQSO where our line of sight just grazes the outflowing high ionization
wind, and misses the low ionization outflow, as suggested by Turnshek et al.
(1994) in the case of PG~0043+039. 

Is the red continuum, excess blue flux in the
high ionization lines, associated absorption, and dense BLR, unique to \zw,
or are they typical UV properties of narrow line quasars? 
No complete UV study of narrow line AGNs is available to answer these
questions, but some hints may be obtained from existing 
observations (see also discussion in Lawrence et al. 1997). 

Baldwin et al. (1996) studied the emission line properties of a
heterogeneous sample of seven $z\sim 2$ quasars with a range of
emission line properties. Some of their objects have strong Fe~II,
Fe~III, and Al~III emission as observed in \zw. Baldwin et
al. analyzed in detail the UV spectrum of Q$0207-398$, where a high
S/N ratio was available, and found excess emission in the blue wing of
the high ionization UV lines, which they interpreted as a high
ionization outflowing component in the BLR, although there was no
direct evidence (through absorption) for such a component, as probably
observed in \zw.

It is also interesting to note that BG92 found that
strong optical Fe~II emitting quasars, which tend to have narrow \hb, also 
tend to have blue excess flux in \hb, and Boroson \& Meyers found that low
ionization BALQSOs, which are generally \zw\ like, have blue excess flux in \ha.
It is not known, however, whether this
property extends to the UV lines as well.

If the UV absorption system in \zw\ is indeed producing the blueshifted
components of the high ionization lines, then the absence of a corresponding
redshifted component indicates that the far side of this
outflow has to be obscured. This may either be due to an extended highly
optically thick gas, such as an accretion disk, or it may result from
absorption within each cloud, if the
clouds column density is large enough (see Ferland et al. 1992 for discussion).

The doublet ratios observed here indicate that
the low ionization lines are thermalized and 
will not be emitted isotropically. However,
the large velocity gradients in the wind may strongly reduce the optical
depth for the high ionization lines in the wind. The observed blueshifted
emission in C~IV and N~V is far too broad to determine  
the observed doublet ratio and cannot be used to determine if the emission
from that gas is optically thick or thin. 
 
\zw\ has particularly weak C~III]~$\lambda~1909$ emission (\S 3.1.2)
but normal Si~III]~$\lambda~1892$ emission, which
we interpret here as evidence for a relatively dense ($\sim 10^{11}$~\cmIII)
BLR, at least for the region which produces the the low ionization lines.
 Baldwin et al. (1988) noted the weakness of C~III]~$\lambda~1909$ 
in four other quasars with
narrow UV lines, and this trend is also apparent in Baldwin et al. (1996)
quasars. The most extreme case is of H$0335-336$ where essentially
no C~III]~$\lambda~1909$ emission was observed (Hartig \& Baldwin 1986). 
This quasar has very narrow lines, and is also a low ionization BALQSO.

Independent evidence for a dense BLR with 
$n_e\gtrsim 10^{11}$~\cmIII\ is provided by the significant EW
and thermalized doublet ratio of the Al~III~$\lambda 1857$ doublet
(\S 3.1.6), and by the significant EW of C~III$^*~\lambda 1176$
(\S 3.1.2).
 Another indication for a dense BLR in \zw\ is provided by
its optical spectrum. The Na~I~$\lambda\lambda 5889.95, 5895.92$ emission
line is rather
strong (Oke \& Lauer 1979; Phillips 1976), which Thompson (1991) and
Korista et al. (1997, their Figure 3g) find can only be produced for 
$n_e\gtrsim 10^{11}$~\cmIII, and a large
column density (required to shield neutral Na~I from ionizing 
radiation at $E>5.14$~eV).

\zw\ has a ``red'' continuum (\S 3.3), which is rather rare among quasars.
Two of the four quasars with 
narrow lines described by Baldwin et al. 1988 also have a $\alpha \approx -1$ UV
spectral slope, but the other two narrow line quasars 
have $\alpha\approx 0$ and $-0.5$, which are typical values.

It thus appears that some of the properties of \zw, in particular the 
relatively weak C~III]~$\lambda~1909$, and the blueshifted excess emission
in the high ionization lines, may be common in narrow line quasars.
A more systematic study of the UV emission of narrow line
AGNs is required to establish their typical emission line properties,
their relation to low ionization BALQSOs, 
and to eventually understand the underlying physics.

\section{Summary} 

A high S/N UV spectrum of \zw\ was obtained with the {\it HST} FOS with
a spectral resolution of $\lambda/\Delta\lambda\sim 1300$ over the
1150-3250~\AA\ spectral region. The following main results are obtained
from our analysis of the UV spectrum of \zw.
\begin{enumerate}

\item The Mg~II~$\lambda 2798$ doublet is partially resolved.
The measured doublet ratio is 1.2/1, consistent with theoretical
predictions that the BLR is highly optically thick to absorption in this
line.

\item The Al~III~$\lambda 1857$ doublet is clearly resolved with
an ``optically thick'' doublet ratio of 1.25/1. The line
optical depth provides an upper limit to the clouds distance from
the ionizing source which is
consistent with the ``standard'' BLR radius. The strength of this
line requires a BLR component with 
$n_e\gtrsim 4\times 10^{11}-4\times 10^{12}$~\cmIII.

\item A weak UV absorption system is detected in N~V, and possibly
C~IV and \lya\ as well, indicating an outflow with a line of sight
velocity of $\sim 1870$\kms\ and
velocity dispersion $\lesssim 300$\kms. 

\item Lines from ions of increasing ionization level show increasing
excess flux on the blue wing, and an increasing shift in the velocity
of the emission line peak,
reaching a maximum blueshift of $\sim 2000$\kms\ for He~II~$\lambda 1640$.
This may indicate an outflowing component in the BLR, visible only in the
approaching direction, where the ionization level increases with velocity.
The highest velocity part of this outflow may produce the observed UV
absorption system.

\item The small C III]$~\lambda 1909$ EW, and the small
C III]$~\lambda 1909/$\lya\ and C III]$~\lambda 1909/$Si III]$~\lambda 1892$
flux ratios indicate that the bulk of the gas in the BLR has a density of 
$10^{11}$~\cmIII, 
about an order of magnitude larger than typically seen in AGNs.

\item Very weak [C III]$~\lambda 1907$ and [Si III]$~\lambda 1883$ emission
may be present, indicating there may be a NLR component with 
$n_e\sim 5\times 10^5$~\cmIII.

\item The C~III$^*~\lambda 1176$ line is clearly detected. Its strength
indicates a BLR component with $n_e\gtrsim 10^{11}$~\cmIII.

\item Prominent Fe~II~UV~191 emission is seen, together with weaker
emission lines at 1294~\AA\ and 1871~\AA. These three features have been 
proposed in the literature as evidence for significant \lya\ pumping
of the 8-10 eV levels of Fe~II.

\item Significant Fe~III emission is present. Fe~III UV~34 and UV~48
multiplets are clearly resolved, and Fe~III UV~1, UV~47, UV~50, and
UV~68 may also be present. The implications of significant Fe~III
emission for the conditions in the BLR needs to be explored.

\item The suggestion by Grandi (1983) that the O~I~$\lambda 1304$ 
emission line is
partly converted to O~I]~$\lambda 1641$ is ruled out in the case of
\zw, indicating that
the bulk of O~I \lyb\ Bowen flouresence occurs in the BLR clouds 
at a depth where $\tau_{Ly\alpha}< 10^9$. The strong
suppression of O~I~$\lambda 1304$ in \zw\ is either due to reddening,
or to Balmer continuum absorption.

\end{enumerate}

The HST spectrum of \zw\ demonstrates the richness and complexity of
quasars emission line spectra. It is not clear whether some of the
emission features are unique to \zw, or whether they are typical of
most AGNs.  Many of the features detected here would be blended and
unidentifiable in typical quasar spectra, creating a deceptively
simple emission spectrum. A close inspection of the Francis et
al. composite (Fig.1) suggests that most of the features detected here
are indeed present in other AGNs. The rich UV emission spectrum of
\zw\ should serve as a useful benchmark for photoionization codes, in
particular for codes attempting to calculate the complex emission
spectrum of Fe~II.  Many weak features remain unidentified, and it is
likely that at least some of these are weak Fe~II features. A reliable
identification of these weak features may follow from detailed
photoionization models.  The high quality optical-UV spectrum of \zw\
obtained in this study, can be used for constructing an Fe~II UV
template, and is available upon request from the authors.

\acknowledgments
Support for this work is provided in part by NASA through Grant No. 
GO-5486.01-93A from the Space Telescope Science Institute, and by
a grant to A. L. from the Israel Science Foundation.
We thank C. L. Joseph for his help in obtaining the ground based spectrum
of \zw, and the referee for a careful reading of the paper and helpful
comments.


\onecolumn
\newpage

\begin{figure}
\plotone{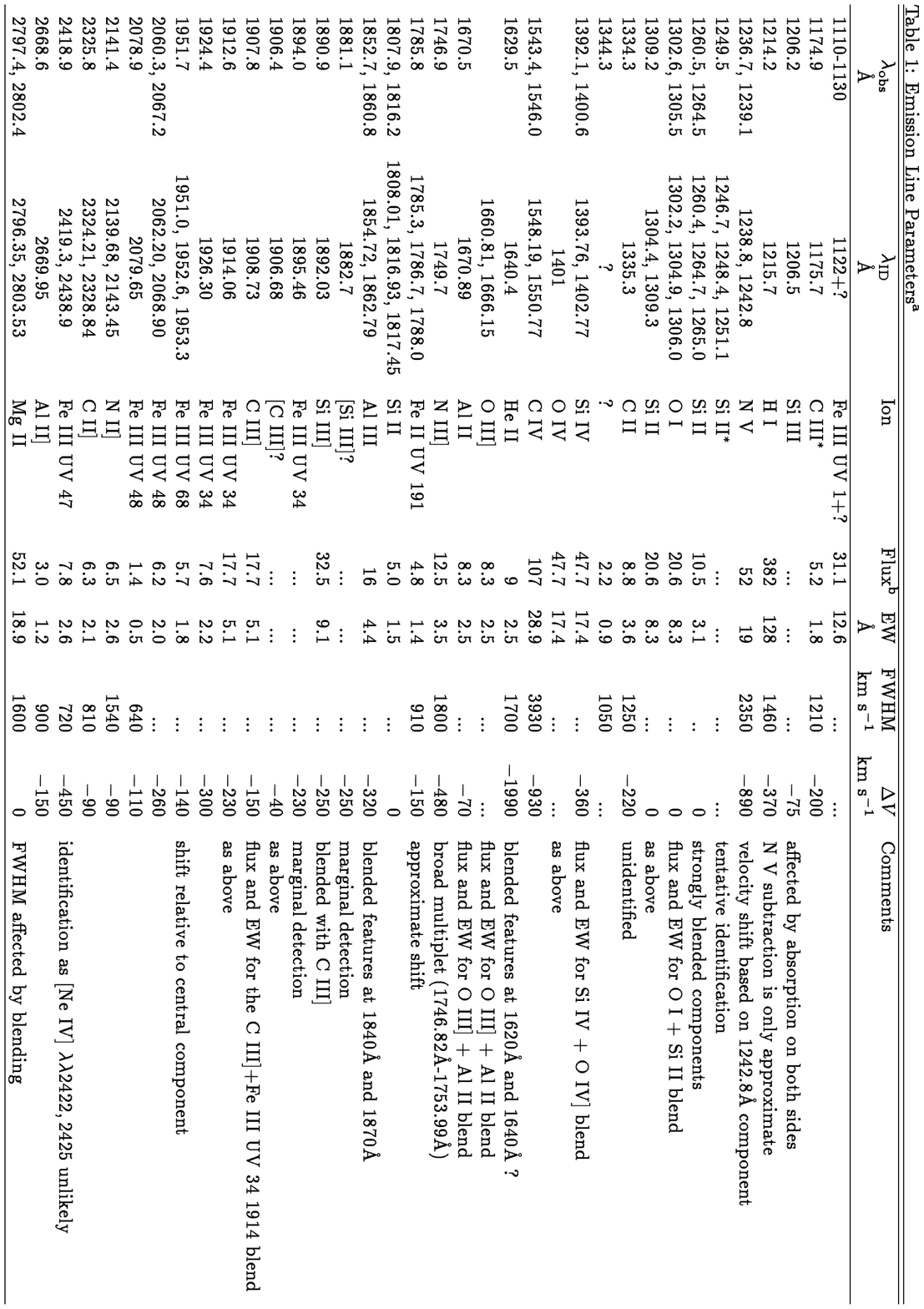}
\end{figure}

\newpage

\begin{figure}
\epsscale{0.8}
\plotone{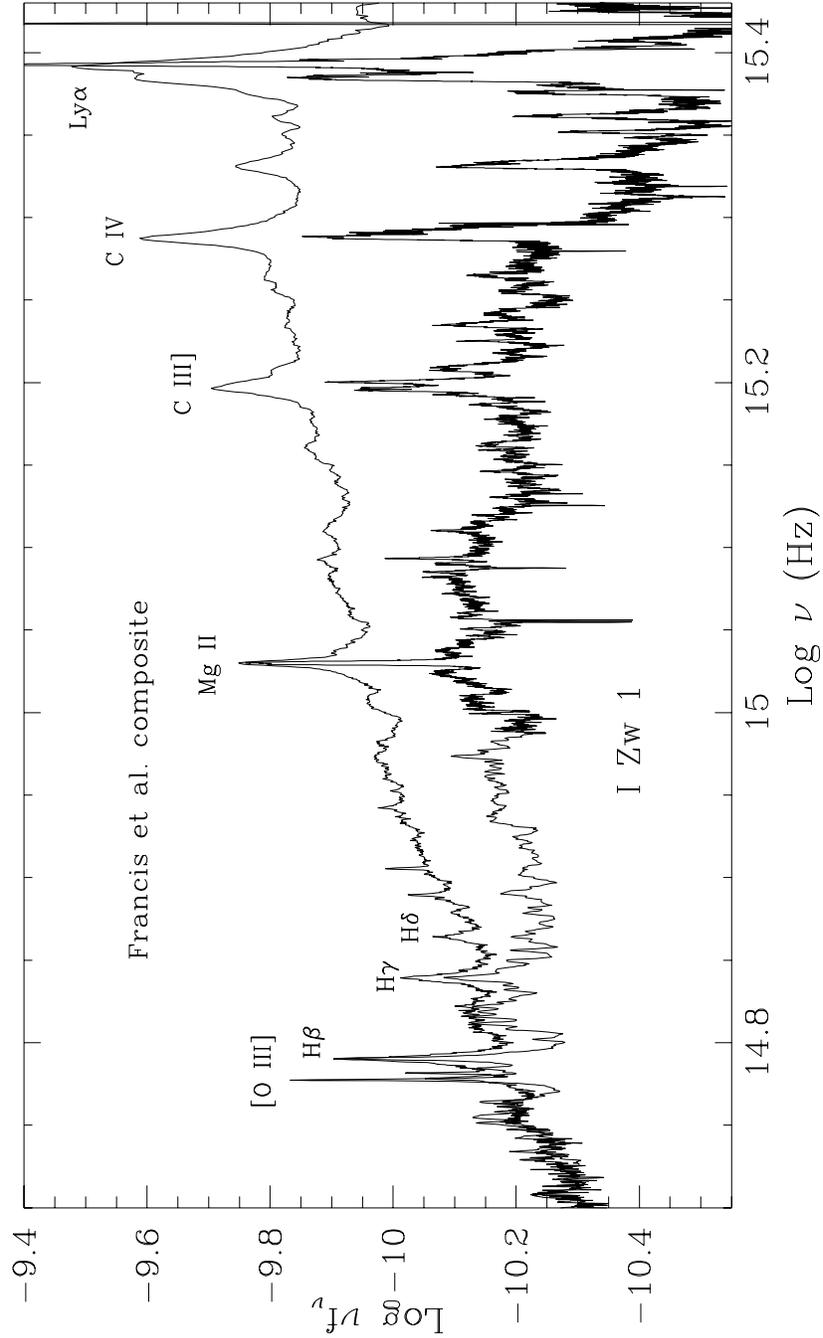}
\caption{ 
The optical-UV emission of \zw, corrected for Galactic reddening (for
E(B$-$V)=0.1), versus the Francis et al.
composite (set to match \zw\ at $\log \nu=14.7$). 
Note that there is practically no wavelength range without
significant emission features. Many of the very weak features in
the Francis et al. composite are clearly resolved in the \zw\ spectrum.
The optical spectrum of \zw\ was available through the courtesy of B. Wills.}
\end{figure}
\normalsize

\newpage
\begin{figure}
\epsscale{0.8}
\plotone{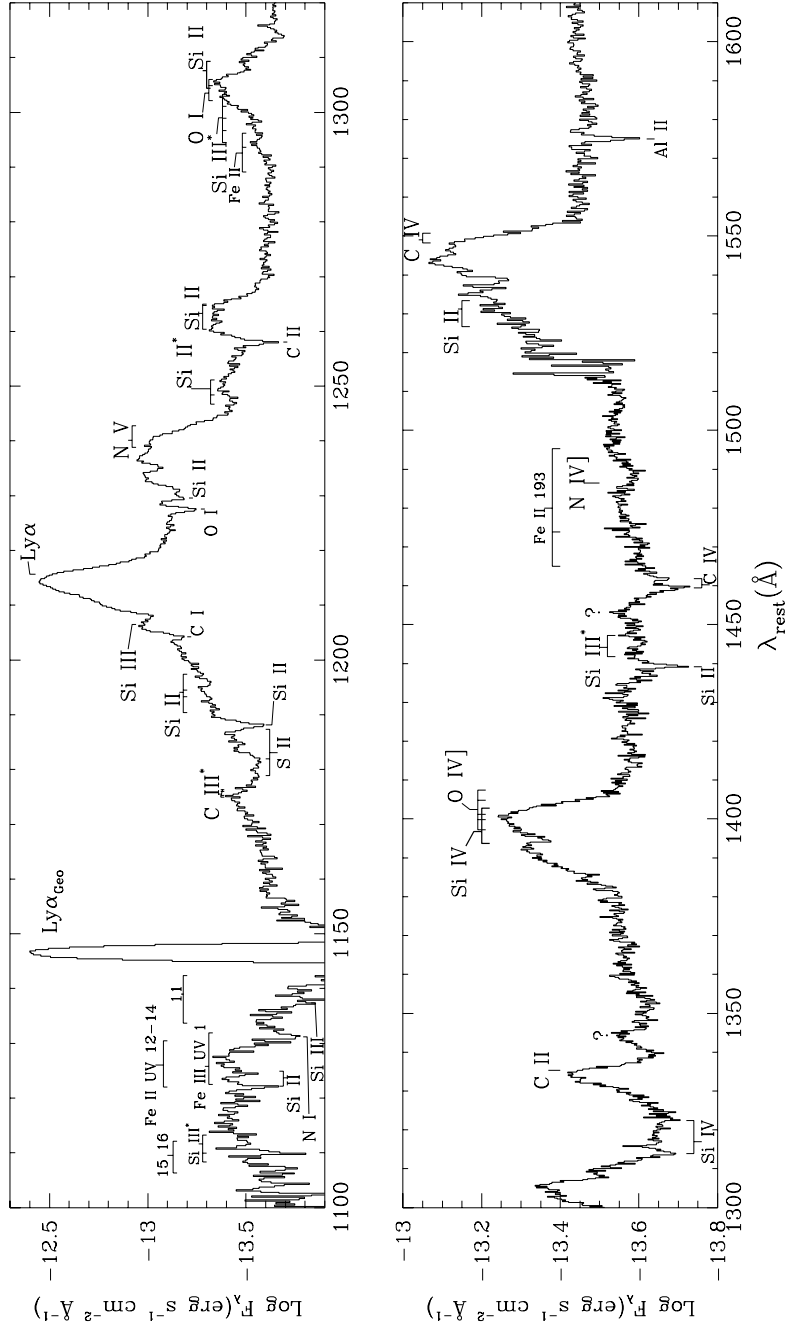}
\caption{
a. Tentative identifications of most of the emission features in the HST FOS 
spectrum of \zw. The lines indicate the expected wavelength for $z=0.0608$.
Only a small number of the Fe~II multiplets which may be present
were marked. Line designations below the spectrum refer to Galactic
absorption features.}
\end{figure}
\normalsize

\newpage
\addtocounter{figure}{-1}
\begin{figure}
\epsscale{0.8}
\plotone{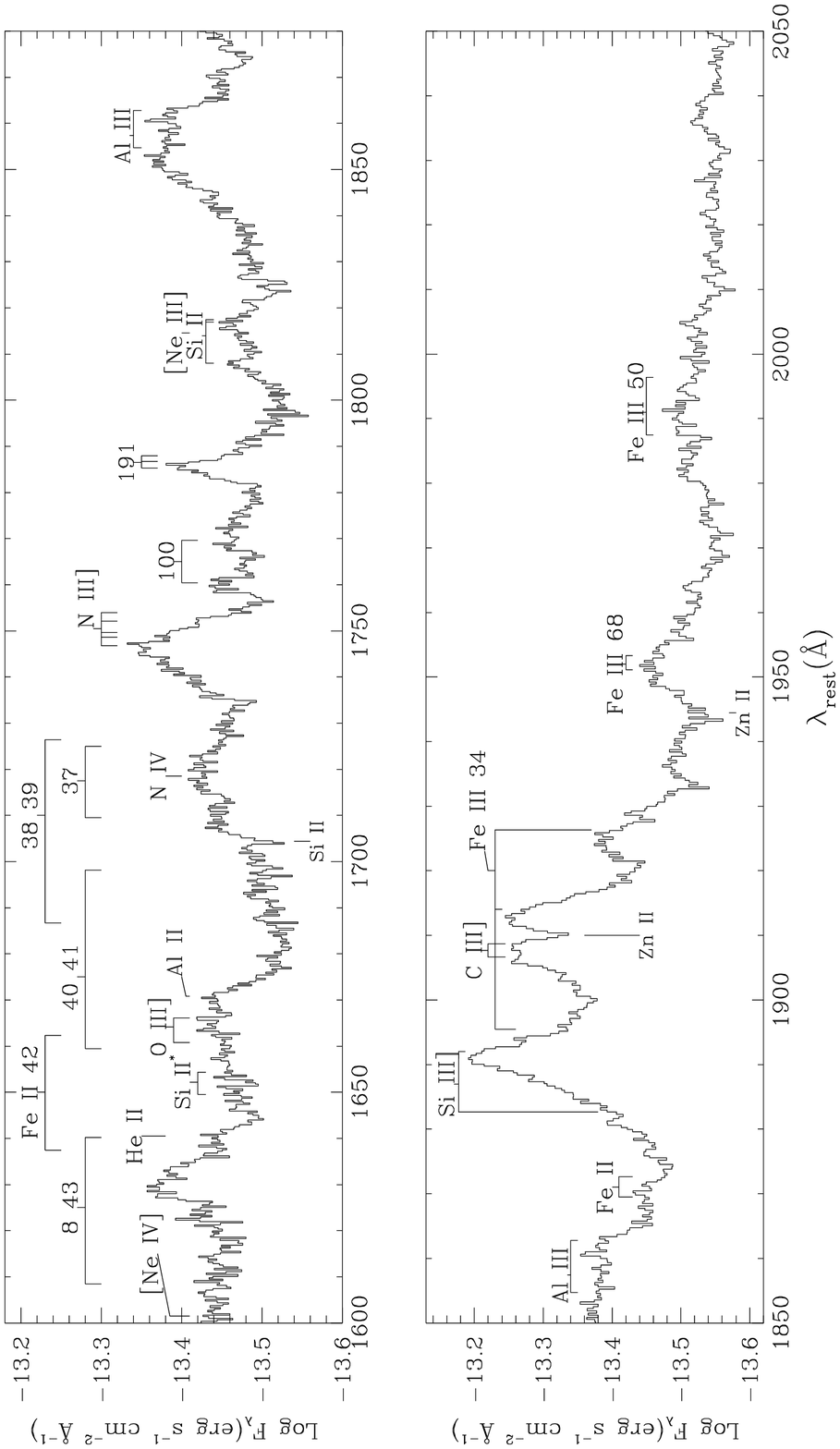}
\caption{
b. As above}
\end{figure}

\newpage
\addtocounter{figure}{-1}
\begin{figure}
\epsscale{0.8}
\plotone{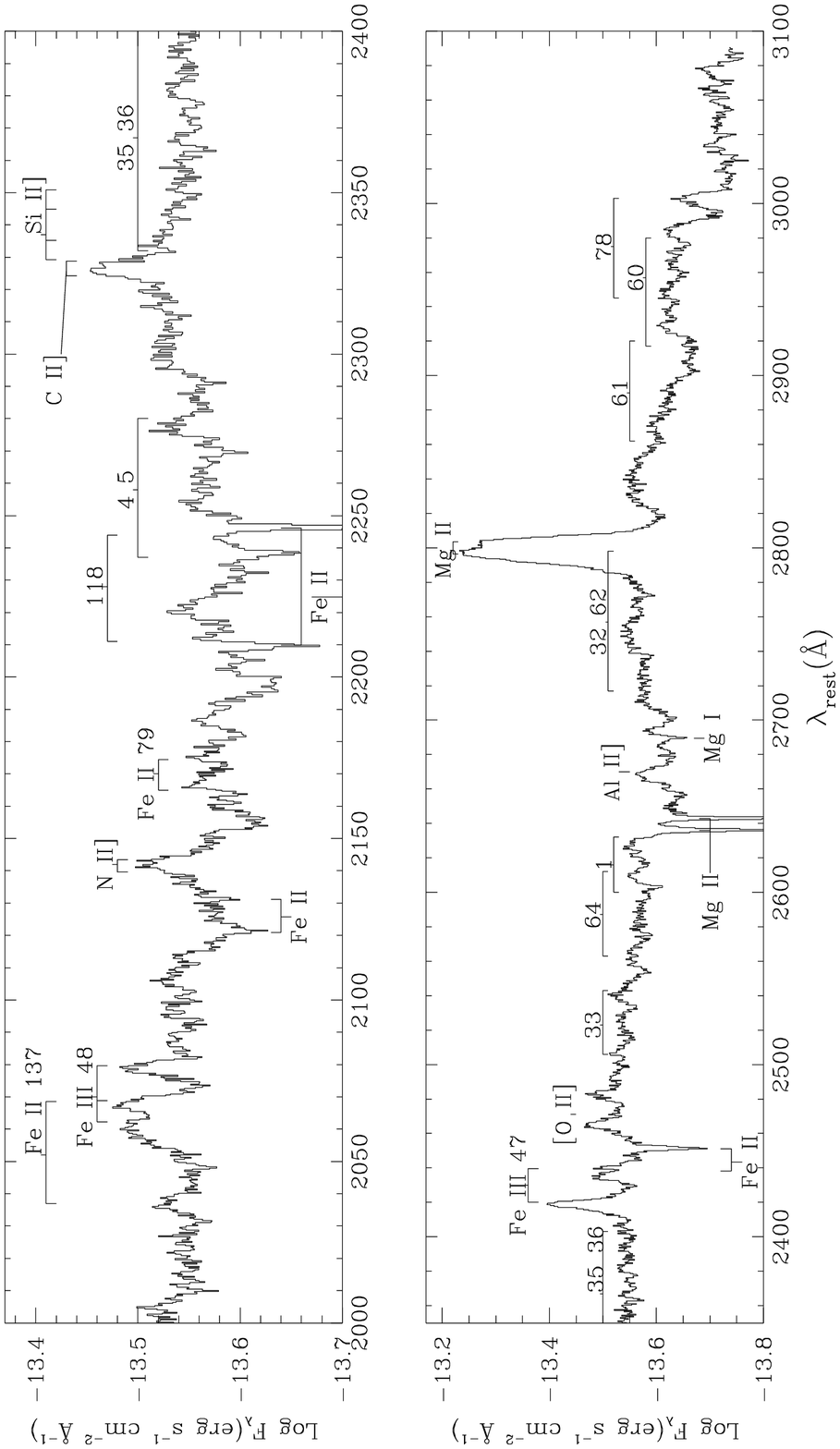}
\caption{
c. As above} 
\end{figure}

\newpage
\addtocounter{figure}{-1}
\begin{figure}
\epsscale{0.8}
\plotone{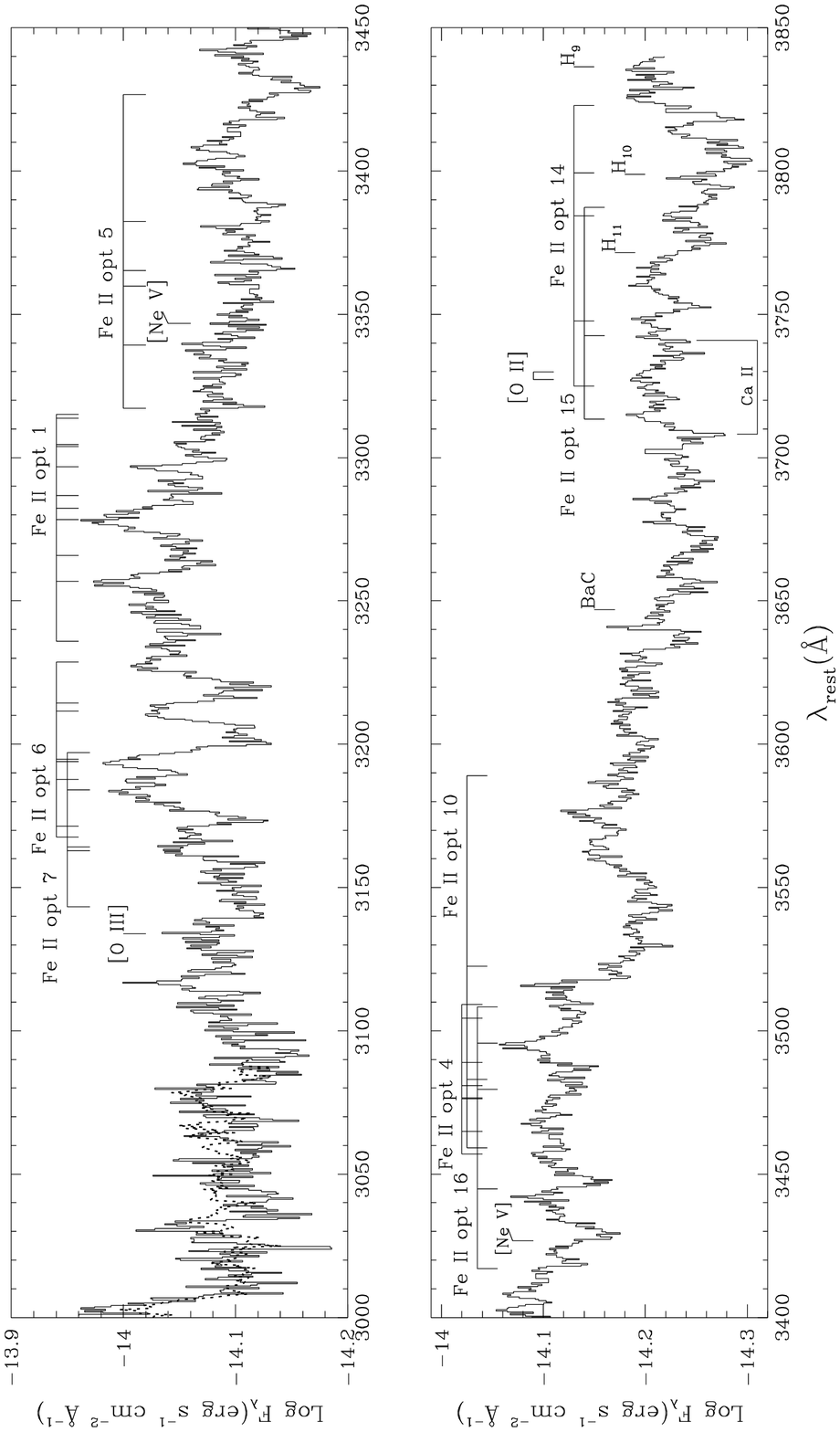}
\caption{
d. The optical spectrum. Note that the absolute flux 
calibration
is not secure. The overlapping HST spectrum is plotted as a dotted line
at $\lambda<3100$~\AA. This spectrum was shifted down by 0.37 in log
flux to match the optical spectrum.}

\end{figure}

\newpage
\begin{figure}
\plotone{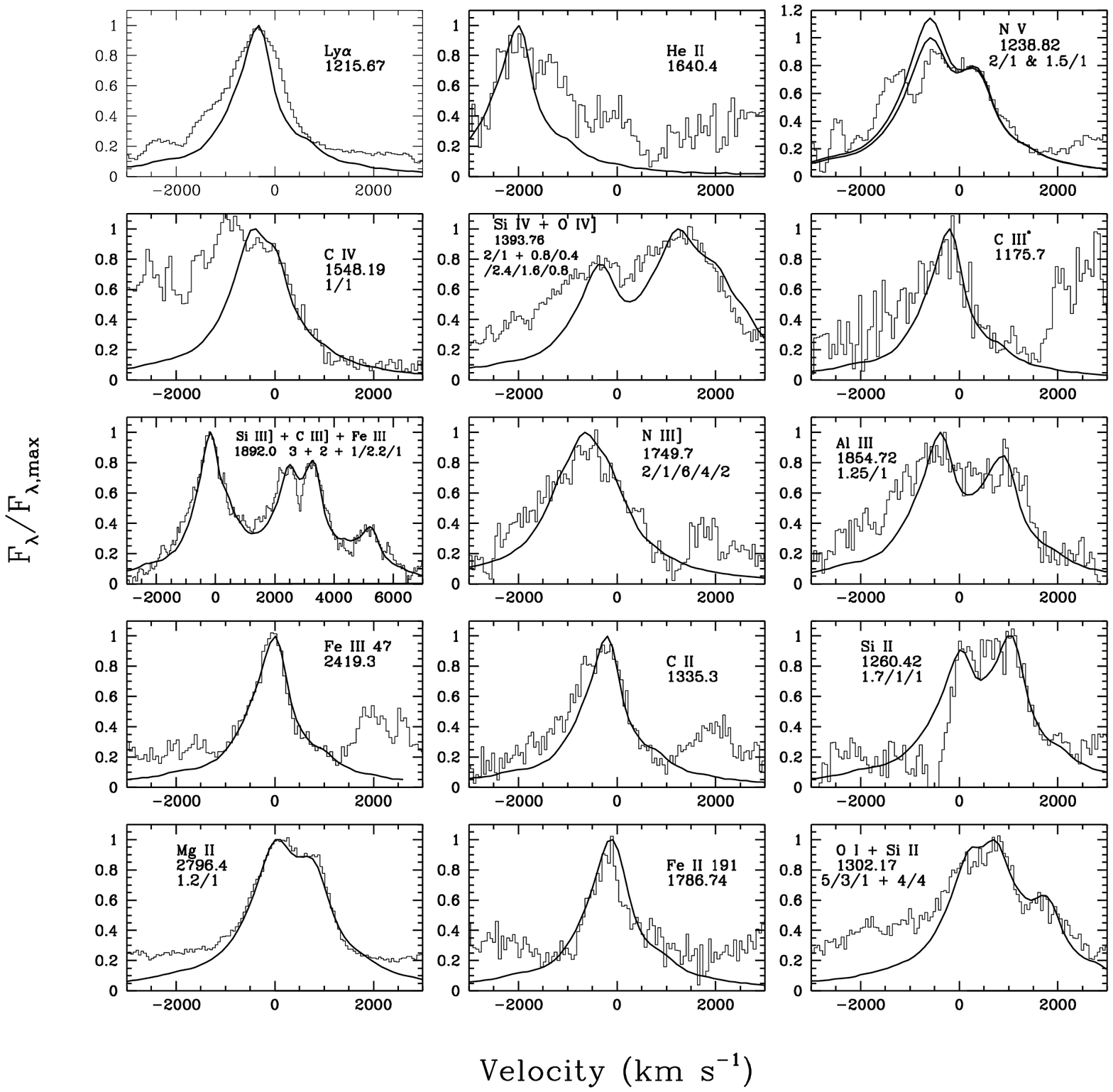}
\caption{
Emission line profiles for selected lines observed in the
spectrum of \zw. The histogram represents the observed profile
and the solid line is the \ha\ profile fit. 
Multiplets are fit with a sum of template components. The 
wavelength 
used to set the velocity scale and the multiplet component flux ratios
are indicated in each panel.
Note the overall bad fit and large velocity shifts of the high ionization 
lines (upper panels), and the much better fit and very small shifts of the 
low ionization lines (lower panels).}

\end{figure}

\newpage
\begin{figure}
\plotone{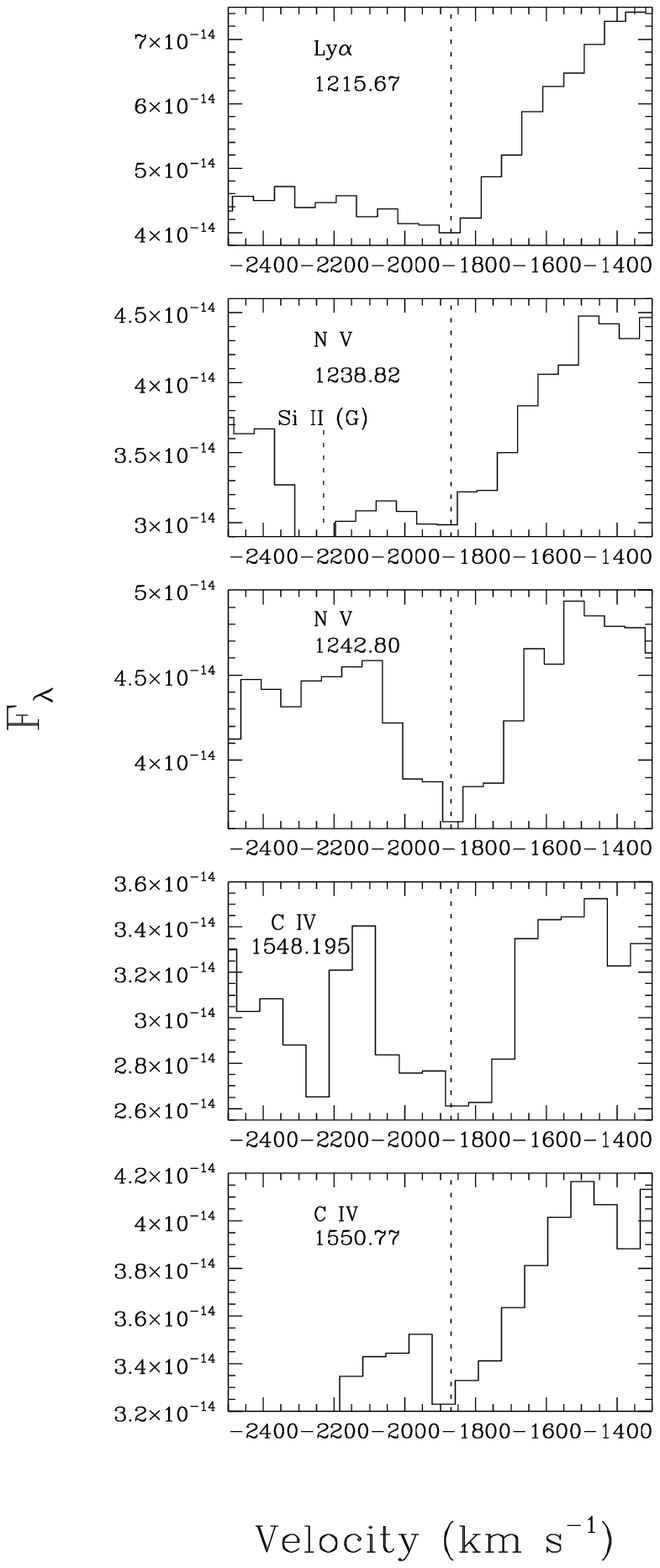}
\caption{
Absorption line profiles of the candidate intrinsic UV absorption line system.
The rest wavelength used to set the velocity scale
is indicated in each panel. The strongest evidence for absorption is seen in
the N~V~$\lambda 1242.80$ profile. The other lines are consistent with
having absorption of similar EW and FWHM at the same velocity shift of
$\sim 1870$~km~s$^{-1}$.}
\end{figure}

\end{document}